%% file: main.tex
\documentclass[letterpaper]{article}
\usepackage[preprint]{aaai2027}
\usepackage[hyphens]{url}
\usepackage{graphicx}
\usepackage{natbib}
\usepackage{algorithm}
\usepackage{algorithmic}
\usepackage{booktabs}
\usepackage{multirow}
\usepackage{array}
\usepackage{amsmath}
\usepackage{amssymb}
\newcommand{\tightlist}{}
\providecommand{\hypertarget}[2]{#2}

\title{MobiInfer: Accelerating On-Device GUI Agents with Workload-Aware Inference}

\author{
Siming Chen\textsuperscript{\rm 1},
Erhu Feng\textsuperscript{\rm 1}\corresponding,
Dahu Feng\textsuperscript{\rm 2},
Yubin Xia\textsuperscript{\rm 1}
}
\affiliations{
\textsuperscript{\rm 1}Institute of Parallel and Distributed Systems,
Shanghai Jiao Tong University\\
\textsuperscript{\rm 2}Department of Precision Instrument,
Tsinghua University\\
Corresponding author: fengerhu1@sjtu.edu.cn
}

\begin{document}

\maketitle

\begin{abstract}
Mobile GUI agents assist users in executing phone tasks through iterative observe--reason--act loops. However, deploying multimodal GUI agents directly on smartphones is challenging due to strict end-to-end latency requirements and significant memory overhead. We observe that GUI-agent workloads exhibit recurring workflows and structured I/O: their prompts contain reusable context, while their outputs follow fixed schemas with phase-dependent predictability. Based on this observation, we present \textbf{MobiInfer}, a workload-aware inference engine for on-device multimodal GUI agents. MobiInfer combines trajectory-aware prefix reuse, phase-indexed vocabulary specialization, and schema-guided speculative decoding to eliminate redundant prefill computation and reduce structured-generation overhead. We implement MobiInfer on top of MNN and evaluate it on AndroidWorld and a real-world dataset using three real Android devices. Across the four model--benchmark settings, MobiInfer achieves average speedups of $2.6\times$ in prefill and $2.5\times$ in decoding, reduces end-to-end task latency by $2.4\times$ on average, and cuts total deployment storage by 30\%, while limiting task-success degradation to less than one percentage point.
\end{abstract}

\input{sections/introduction}
\input{sections/background}
\input{sections/method}
\input{sections/experiments}
\input{sections/related_work}
\input{sections/conclusion}

\bibliography{references}

\end{document}

%% file: sections/introduction.tex
\hypertarget{introduction}{%
\section{Introduction}\label{introduction}}

Multimodal GUI agents turn a user goal and the current smartphone screen into
an executable action, then repeat this observe--reason--act loop until the task
is complete. Even when applications expose APIs, GUI agents remain necessary for
the long tail of mobile tasks whose functionality is available only through
visual interfaces, whose state is visible only on screen, or whose completion
requires composing actions across multiple applications. Running this loop on
the phone is attractive for latency, privacy, and connectivity reasons, but it
makes every model invocation part of an interactive control path. The practical
bottleneck is not only model size: mobile runtimes repeatedly prefill long
multimodal prompts and decode structured actions under tight compute,
memory-bandwidth, and RAM budgets.

\begin{figure}[t]
\centering
\includegraphics[width=\columnwidth]{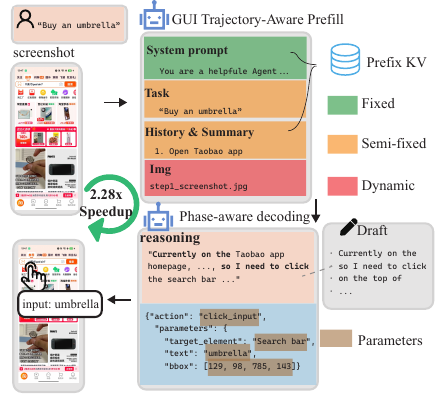}
\caption{MobiInfer exploits the structure of a GUI-agent trajectory.
GUI trajectory-aware prefill groups fixed and semi-fixed prompt fields into a
reusable prefix KV, leaving the current screenshot and step-specific context as
the dynamic suffix. Phase-aware decoding drafts predictable reasoning and schema
regions while retaining target-model generation for execution-critical action
parameters.}
\label{fig:mobiinfer-overview}
\end{figure}

General-purpose on-device inference runtimes are not designed around this
trajectory structure. When a GUI agent is executed step by step, the runtime
often receives each model call as a separate request. Stable instructions can be
prepared offline, task context can be reused within a trajectory, and only the
current screenshot and recent UI state must be recomputed at every step. Yet
these fields are often encoded together from scratch. Outputs are likewise
decoded through a full-vocabulary projection despite moving through phases with
different predictability and execution risk: reasoning errors may be
recoverable, whereas an incorrect action label or coordinate can control the
wrong UI element.

We present \textbf{MobiInfer}, an inference engine built around this diagnosis.
\textbf{GUI trajectory-aware prefill} canonicalizes prompt fields by lifetime,
materializes compatible fixed and semi-fixed KV states, and evaluates only the
dynamic suffix. \textbf{Phase-indexed vocabulary specialization} selects a
compact output head from the committed generation phase. \textbf{Schema-guided
speculative decoding} drafts only predictable structure and reasoning; it stops
before action labels, coordinates, and other execution-critical values, which
the target model generates and verifies. Together, these mechanisms align
prefill reuse and decoding acceleration with the structure and risk profile of
each GUI-agent trajectory without retraining the agent model or changing its
external action interface.
Figure~\ref{fig:mobiinfer-overview} summarizes how these mechanisms map onto
the repeated agent loop.

We implement MobiInfer over MNN and evaluate it on AndroidWorld and a real-world
dataset using three real Android devices. Against the deployment MNN baseline,
MobiInfer achieves average speedups of $2.6\times$ in prefill, $2.5\times$ in
decoding, and $2.4\times$ in end-to-end task latency while keeping
task-success degradation below one percentage point. Moreover,
the peak memory footprint decreases by 31.8\%, and total deployment storage
decreases by 30.0\%.
We make three contributions:

\begin{itemize}
\tightlist
\item We identify GUI-agent inference as a long-lived trajectory with reusable
prompt prefixes and phase-structured, risk-sensitive outputs.
\item We design a workload-aware runtime that jointly accelerates prefill and
structured decoding without retraining the agent model.
\item We evaluate MobiInfer on Android devices with matched mobile
baselines, measuring task latency, stage throughput, task success, memory, and
deployment storage.
\end{itemize}

%% file: sections/background.tex
\hypertarget{motivation-and-characterization}{%
\section{Motivation and Characterization}\label{motivation-and-characterization}}

We quantify two workload regularities exploited by MobiInfer: reusable prompt
regions and phase-structured outputs.

\subsection{Reusable Prompt Regions}

Across a GUI-agent trajectory, large prompt spans remain unchanged across model
calls: instructions, tool descriptions, output schemas, the user goal, and
committed history often reappear as the screen observation changes. This creates
substantial repeated prefill work, but it is reusable only when the repeated
tokens form a contiguous autoregressive prefix. If a prompt template interleaves
observation-dependent fields with stable fields, a small screen change can
invalidate the KV states of all following tokens.

\subsection{Phase-Structured Outputs}

GUI-agent outputs are structured but not uniformly predictable. A response moves
through JSON structure, reasoning, an action label, and typed arguments such as
text or coordinates. These phases differ in both token space and execution risk:
schema tokens are narrow and repetitive, reasoning is broader, and actions or
coordinates depend on the current screen and directly control the device.

\begin{figure}[t]
\centering
\includegraphics[width=\columnwidth]{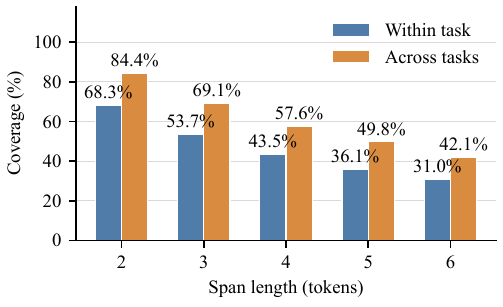}
\caption{Prevalence of repeated consecutive token spans in agent outputs.
Across-task coverage is the fraction of held-out windows whose span occurs in
at least 20 distinct collected real-world GUI-agent traces.}
\label{fig:span-repetition}
\end{figure}

\begin{figure}[t]
\centering
\includegraphics[width=\columnwidth]{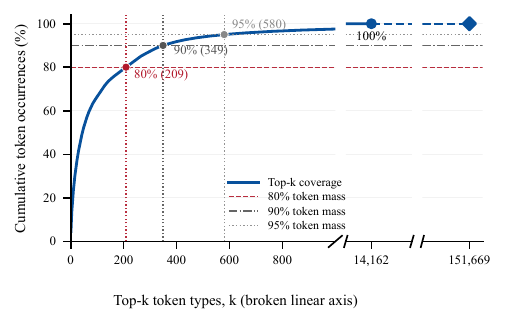}
\caption{Token-frequency concentration in more than 7,000 collected real-world
GUI-agent traces.
Token types are ranked by occurrence frequency. Only 209, 349, and 580 types
are required to cover 80\%, 90\%, and 95\% of all token occurrences,
respectively. Of the full 151,669-entry tokenizer vocabulary, 14,162 types
occur in the dataset. The broken linear axis preserves the high-frequency head
while marking the observed-vocabulary endpoint and the full tokenizer size.}
\label{fig:token-concentration}
\end{figure}

Figure~\ref{fig:span-repetition} first measures repeated consecutive token spans
across task-disjoint outputs. Figure~\ref{fig:token-concentration} then shows
that GUI-agent outputs use only a small fraction of the tokenizer vocabulary,
with a few hundred token types covering most occurrences. This concentration
suggests that many decoding steps do not require full-vocabulary projection.

Longer spans remain recurrent across task-disjoint outputs: 69.1\% of 3-token
windows and 57.6\% of 4-token windows occur in at least 20 distinct collected
real-world GUI-agent traces. Together, token concentration and span recurrence
motivate MobiInfer's output-side design: phase-indexed projection reduces
large-head cost, while drafting is enabled only for predictable, non-critical
phases.

%% file: sections/method.tex
\hypertarget{method}{%
\section{Method}\label{method}}

MobiInfer combines three runtime mechanisms: GUI trajectory-aware prefill,
phase-indexed vocabulary specialization, and schema-guided speculative decoding.
All fast paths preserve the original token interface and commit drafted tokens
only after target-model verification.

\subsection{GUI Trajectory-Aware Prefill}
\label{kv-cache-precomputation-and-restoration}

\paragraph{Prompt canonicalization.}
Let a rendered prompt contain fixed fields $P_f$, semi-fixed fields $P_s$, and
dynamic fields $P_d$. MobiInfer constructs
\begin{equation}
P_{\mathrm{can}} = P_f \mathbin{\|} P_s \mathbin{\|} P_d,
\end{equation}
where $\|$ denotes concatenation with explicit template delimiters. Order is
preserved within each region. System instructions, tool definitions, and the
output schema belong to $P_f$; the user goal and stable task context belong to
$P_s$; screenshots, visual tokens, OCR, and recent UI state belong to $P_d$.
Any field that cannot be classified safely is assigned to $P_d$.
This transformation makes $P_f \| P_s$ a reusable autoregressive prefix.

\paragraph{Prefix materialization and restoration.}
For a reusable prefix $P_r$, MobiInfer runs the target model once and stores the
layer-wise key and value tensors. 
Globally fixed states are prepared offline, stored on disk as part of the
deployment package, and paired with the exact model, tokenizer, and canonical
template used at deployment. Semi-fixed states are materialized after the user
goal is known and reused only within that trajectory. The cached-prefix package
stores enough metadata to validate token, position, model, and backend
compatibility before restoration.

At runtime, the engine loads and restores the fixed cached prefix and, when
available, the semi-fixed state. It sets the cached sequence length as the
past-context length and begins $P_d$ at the immediately following position. A changed user goal
creates a new semi-fixed state rather than reusing the preceding task's state.
Under identical tokens, positions, weights, and runtime kernels, restoration is
functionally equivalent to recomputing the prefix.

Within a running task, MobiInfer also supports delta prefill. If the current
canonical prompt extends the previous prompt, the engine retains the active KV
state and processes only newly appended tokens.

\subsection{Phase-Indexed Vocabulary Specialization}
\label{vocabulary-and-lm-head-specialization}

MobiInfer reduces vocabulary cost in two stages: (1) \emph{global row pruning}
and (2) \emph{phase-indexed active-head selection}. First, it performs global
row pruning on both the token embedding and LM-head matrices. As shown in
Figure~\ref{fig:token-concentration}, our collected real-world GUI-agent data
uses only 14,162 of the 151,669 tokenizer entries. MobiInfer keeps these
observed token rows, plus schema-required tokens, special tokens, and the
complete base-byte alphabet, and removes the remaining rows from both matrices.
The byte alphabet, consisting of 256 tokens, provides coverage fallback, so with
BPE-style tokenization, text can still be emitted through byte tokens even when
a larger merged token is pruned.

Second, MobiInfer selects a smaller active head for each output phase. It
represents the output schema as an incremental automaton $\mathcal{A}$ whose
state $q$ indicates whether the model is generating structure, reasoning, an
action label, coordinates, another typed parameter, or completion. Using
collected outputs labeled by this automaton, MobiInfer builds a phase vocabulary
\begin{equation}
V_q = V_{\mathrm{schema}}(q) \cup V_{\mathrm{freq}}(q)
      \cup V_{\mathrm{byte}} \cup V_{\mathrm{special}}.
\end{equation}
$V_{\mathrm{schema}}(q)$ contains tokens required by legal schema transitions,
$V_{\mathrm{freq}}(q)$ contains frequent tokens from collected output segments
assigned to phase $q$ by the automaton,
$V_{\mathrm{byte}}$ contains the 256 byte tokens, and $V_{\mathrm{special}}$
contains required control and termination tokens. Structure, action, and
coordinate phases therefore use narrow vocabularies, while reasoning keeps a
broader set of frequent tokens.

At runtime, the current automaton state $q$ selects the phase head for hidden
state $h$ and output matrix $W$:
\begin{equation}
W_q = W[V_q,:], \qquad \ell_q = W_q h.
\end{equation}
The schema mask is applied inside $V_q$, and the selected row is mapped back to
the original tokenizer ID before the token is committed. Thus MobiInfer keeps
the external token interface unchanged while avoiding full-vocabulary projection
for phases that need only a small part of the vocabulary.

\begin{algorithm}[ht]
\caption{Schema-guided decoding in MobiInfer}
\label{alg:structure-aware-speculation}
\small
\begin{algorithmic}[1]
\REQUIRE Target model $M$, prompt $x$, automaton $\mathcal{A}$, n-gram table $G$
\REQUIRE Byte-complete phase heads $\{W_q\}$, protected states $\mathcal{S}_{\mathrm{prot}}$
\ENSURE Structured agent output $y$
\STATE $K \gets \textsc{PrefillOrRestore}(M,x)$
\STATE $y \gets \emptyset$; $q \gets \mathcal{A}.q_0$
\WHILE{$q \neq q_{\mathrm{done}}$}
  \STATE $H \gets \textsc{SelectHead}(q,\{W_q\})$
  \IF{$q \in \mathcal{S}_{\mathrm{prot}}$}
    \STATE $(v,K) \gets \textsc{DecodeOne}(M,H,K)$
    \STATE $(y,q) \gets \textsc{CommitOne}(y,q,v,\mathcal{A})$
  \ELSE
    \STATE $d \gets \textsc{Draft}(q,G,\mathcal{A},y)$
    \STATE $d \gets \textsc{SafePrefix}(d,q,\mathcal{A},\mathcal{S}_{\mathrm{prot}})$
    \IF{$d=\emptyset$}
      \STATE $(v,K) \gets \textsc{DecodeOne}(M,H,K)$
      \STATE $(y,q) \gets \textsc{CommitOne}(y,q,v,\mathcal{A})$
    \ELSE
      \STATE $(z,K_d) \gets \textsc{Verify}(M,K,d,\mathcal{A})$
      \STATE $m \gets \textsc{LongestMatch}(d,z)$
      \STATE $(y,K,q) \gets \textsc{Commit}(y,K,K_d,d,z,m,\mathcal{A})$
    \ENDIF
  \ENDIF
\ENDWHILE
\STATE \textbf{return} $\textsc{ParseAgentOutput}(y)$
\end{algorithmic}
\end{algorithm}

\subsection{Schema-Guided Speculative Decoding}
\label{structure-aware-speculative-decoding}

The same automaton controls where MobiInfer speculates. MobiInfer separates
output phases into speculative states and protected states:
\begin{align}
\mathcal{S}_{\mathrm{spec}}
&= \{q_{\mathrm{struct}}, q_{\mathrm{reason}}\}, \\
\mathcal{S}_{\mathrm{prot}}
&= \{q_{\mathrm{action}}, q_{\mathrm{coord}}, q_{\mathrm{param}}\}.
\end{align}
MobiInfer uses a learning-free
n-gram drafting algorithm. To construct $G$, MobiInfer tokenizes the collected
real-world agent-output sequences, assigns each token span to an automaton
phase, and stores frequent next-token continuations for reasoning states.
In addition to this offline table, MobiInfer maintains a trajectory-local
n-gram cache from verified agent outputs produced earlier in the current task.
Task-local entries are preferred over offline entries when both match the same
suffix, and the cache is discarded at the end of the trajectory.
During decoding, structure states propose fixed schema literals, such as field
names and delimiters, as draft candidates. Reasoning states query $G$ with the
longest suffix of the committed output $y$ to obtain a short candidate
continuation. Protected states use target-only
decoding; action labels and coordinates are always protected. If a draft would
cross into a protected state, MobiInfer simulates it on a copy of $\mathcal{A}$
and keeps only the longest safe prefix.
For example, if a draft contains a reasoning suffix followed by the opening
token of an action value, \textsc{SafePrefix} retains only the reasoning suffix
and leaves the entire action value to target-only decoding.
Algorithm~\ref{alg:structure-aware-speculation} shows the joint decoding path.

Verification evaluates the draft with the target model in one batched pass and
commits only the longest matching prefix. Under greedy decoding, speculation
preserves the output of the phase-specialized target policy; vocabulary
specialization itself is approximate and may differ from full-vocabulary
decoding. When no safe draft exists, MobiInfer falls back to one-token decoding.
When outputs contain little recurring structure, the n-gram drafter may provide
no useful continuation; MobiInfer then uses this fallback without changing the
phase-specialized policy's output.

%% file: sections/experiments.tex
\section{Experiments}
\label{experiments}

\subsection{Experimental Setup}

We evaluate two multimodal GUI-agent models, MAI-UI-2B~\cite{zhou2025maiui} and
GELab-Zero-4B~\cite{stepfun2025gelabzero}. Both models are quantized with the
AutoRound tool~\cite{intel2023autoround} using INT8 GPTQ-based post-training
quantization~\cite{frantar2023gptq}, and are deployed through the MNN LLM
runtime.
Experiments run on three Android phones: OnePlus 13, OPPO Find X8, and Redmi
K50, covering a current Qualcomm flagship, a current MediaTek flagship, and an
older MediaTek platform. The workloads cover two task sources: AndroidWorld, a
fully interactive benchmark with 116 dynamically parameterized task templates
over 20 real applications~\cite{rawles2025androidworld}, and our RealWorld, a set
of 111 measured phone tasks grouped into Navigation, System Control, Search,
and Commerce. Additional hardware details and the complete task inventory are
provided in the supplementary material.
The primary baseline is the deployment MNN baseline: the original MNN execution
path with the original agent prompt template, using the same phone, model
weights, tokenizer, quantization configuration, and action schema as MobiInfer.
 MobiInfer
uses a canonical prompt layout to expose
contiguous reusable prefixes; this reordering does not change prompt length, and
we evaluate task success to ensure that the optimized path preserves agent
quality. We report efficiency, resource footprint, and task quality.

Both systems start from the same pretrained checkpoint and task inputs, while
MobiInfer applies its vocabulary and tokenizer transformation; all quantization,
decoding, device, and task-seed settings are matched, and cache construction,
verification, and auxiliary processing are included in its runtime measurements.

\subsection{Main Results and Task Quality}

\begin{figure}[ht]
\centering
\includegraphics[width=\columnwidth]{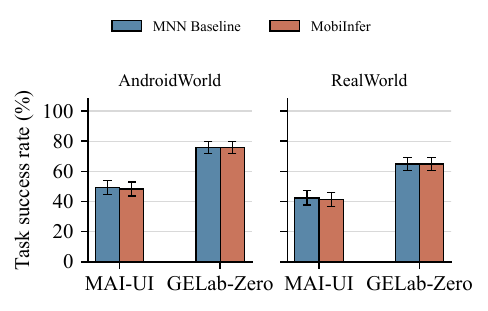}
\caption{Task success rate of the deployment MNN baseline and MobiInfer
across two GUI-agent models and two benchmarks.}
\label{fig:task-success}
\end{figure}

\begin{table*}[!t]
\centering
\small
\setlength{\tabcolsep}{2.0pt}

\begin{tabular}{@{}lllrrrrrrrrr@{}}
\toprule
& & & \multicolumn{3}{c}{E2E latency (s)} &
\multicolumn{3}{c}{Prefill speed} & \multicolumn{3}{c}{Decode speed} \\
\cmidrule(lr){4-6}\cmidrule(lr){7-9}\cmidrule(lr){10-12}
Model & Benchmark & Task category & MNN & MobiInfer & Speedup &
MNN & MobiInfer & Speedup & MNN & MobiInfer & Speedup \\
\midrule
\multirow{9}{*}{MAI-UI-2B} & \multirow{4}{*}{AndroidWorld} & Easy & 49.25 & 26.71 & 1.84$\times$ & 133.37 & 228.93 & 1.72$\times$ & 20.23 & 53.71 & 2.65$\times$ \\
 &  & Medium & 96.23 & 43.46 & 2.21$\times$ & 140.93 & 353.57 & 2.51$\times$ & 19.00 & 47.84 & 2.52$\times$ \\
 &  & Hard & 283.28 & 88.72 & 3.19$\times$ & 112.83 & 520.88 & 4.62$\times$ & 15.52 & 43.69 & 2.82$\times$ \\
 &  & \textbf{Overall} & 85.81 & 39.89 & \textbf{2.15$\times$} & 142.86 & 340.57 & \textbf{2.38$\times$} & 19.24 & 48.12 & \textbf{2.50$\times$} \\
\cmidrule(lr){2-12}
 & \multirow{5}{*}{RealWorld} & Navigation & 21.30 & 12.19 & 1.75$\times$ & 120.18 & 181.32 & 1.51$\times$ & 21.29 & 59.66 & 2.80$\times$ \\
 &  & System Control & 70.09 & 34.40 & 2.04$\times$ & 133.46 & 264.75 & 1.98$\times$ & 20.24 & 52.85 & 2.61$\times$ \\
 &  & Search & 86.57 & 39.87 & 2.17$\times$ & 138.13 & 314.11 & 2.27$\times$ & 19.87 & 50.45 & 2.54$\times$ \\
 &  & Commerce & 832.71 & 185.78 & 4.48$\times$ & 96.53 & 675.26 & 7.00$\times$ & 13.91 & 39.57 & 2.84$\times$ \\
 &  & \textbf{Overall} & 118.32 & 50.83 & \textbf{2.33$\times$} & 138.90 & 364.25 & \textbf{2.62$\times$} & 19.22 & 48.19 & \textbf{2.51$\times$} \\
\midrule
\multirow{9}{*}{GELab-Zero-4B} & \multirow{4}{*}{AndroidWorld} & Easy & 78.53 & 38.25 & 2.05$\times$ & 82.32 & 150.33 & 1.83$\times$ & 12.52 & 33.33 & 2.66$\times$ \\
 &  & Medium & 153.73 & 62.06 & 2.48$\times$ & 86.99 & 232.19 & 2.67$\times$ & 11.76 & 29.68 & 2.52$\times$ \\
 &  & Hard & 456.08 & 126.71 & 3.60$\times$ & 69.65 & 342.05 & 4.91$\times$ & 9.61 & 27.11 & 2.82$\times$ \\
 &  & \textbf{Overall} & 136.99 & 56.91 & \textbf{2.41$\times$} & 88.18 & 223.65 & \textbf{2.54$\times$} & 11.91 & 29.86 & \textbf{2.51$\times$} \\
\cmidrule(lr){2-12}
 & \multirow{5}{*}{RealWorld} & Navigation & 34.02 & 17.24 & 1.97$\times$ & 74.19 & 119.07 & 1.61$\times$ & 13.20 & 36.99 & 2.80$\times$ \\
 &  & System Control & 111.99 & 49.09 & 2.28$\times$ & 82.38 & 173.86 & 2.11$\times$ & 12.55 & 32.77 & 2.61$\times$ \\
 &  & Search & 138.39 & 56.76 & 2.44$\times$ & 85.27 & 206.27 & 2.42$\times$ & 12.32 & 31.28 & 2.54$\times$ \\
 &  & Commerce & 1276.67 & 237.33 & 5.38$\times$ & 51.64 & 466.39 & 9.03$\times$ & 8.79 & 24.75 & 2.82$\times$ \\
 &  & \textbf{Overall} & 189.38 & 72.43 & \textbf{2.61$\times$} & 85.74 & 239.20 & \textbf{2.79$\times$} & 11.92 & 29.88 & \textbf{2.51$\times$} \\
\bottomrule
\end{tabular}

\caption{Aggregated end-to-end latency, prefill throughput, and decode
throughput across two benchmarks and two models. Latency is the mean per-task
latency in seconds. Prefill and decode throughput are computed as total
logical tokens divided by total time in the corresponding stage, and averaged
over three independent repetitions. MNN denotes the deployment baseline using
the original agent prompt template. Throughput speedup and latency speedup are
computed relative to the corresponding MNN values.}
\label{tab:main-results}
\end{table*}

\begin{figure*}[!t]
\centering
\includegraphics[width=\textwidth]{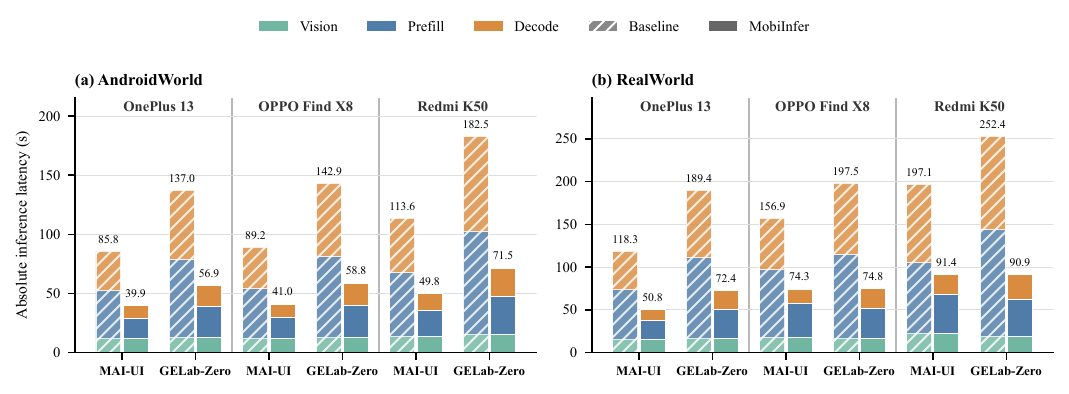}
\caption{Absolute inference-latency breakdown on AndroidWorld and RealWorld
tasks. Each device--model group compares the baseline with MobiInfer across
vision processing, prefill, decoding, and other runtime overhead.}
\label{fig:latency-breakdown}
\end{figure*}

The main results report efficiency together with the task-quality constraint.
Table~\ref{tab:main-results} gives aggregate prefill and decode throughput by
model, benchmark, and task category, with speedups computed against the
deployment baseline for the same model and workload. Figure~\ref{fig:latency-breakdown}
adds absolute stage-level latency, showing that MobiInfer reduces the targeted
prefill and decoding costs while leaving vision cost visible.
The trend is consistent across both GUI-agent models. AndroidWorld shows larger
prefill and decode throughput gains than RealWorld, because its templated tasks
expose more repeated instruction, schema, and history structure across steps.
RealWorld still benefits across all four categories, but its measured phone
tasks contain more diverse observations and action arguments, leaving a larger
fraction of each call on the conservative target-model path. This difference is
useful: it shows that MobiInfer's gains track the amount of reusable and
phase-predictable work rather than a benchmark-specific artifact.

Figure~\ref{fig:task-success} reports the paired task-quality comparison, with
the detailed model--benchmark results provided in the supplementary material.
MobiInfer reduces overall task success from 58.1\% to 57.7\%, a 0.4
percentage-point drop, with the largest drop being 0.9 percentage points across
the four model--benchmark settings. We therefore interpret the latency and
throughput improvements under a task-quality constraint rather than as
unconstrained decoding acceleration. This matters for GUI agents because
format-valid outputs are not sufficient: the generated action must still select
the correct operation and grounded target.

\subsection{Quality Preservation under Prompt and Vocabulary Transformations}
\label{sec:quality-preservation}

We evaluate prompt canonicalization and vocabulary specialization separately
and jointly using four variants: the original prompt with the full vocabulary,
each transformation alone, and the full MobiInfer configuration. All variants
use greedy decoding and identical model weights, tokenizer, quantization,
observations, task states, and action schema. For each step, outputs are
generated from the same screenshot and task state before execution, providing
a paired comparison that is independent of trajectory differences.

We measure target grounding with mean intersection-over-union (IoU) and the
fraction of actions with IoU $\geq 0.5$, after converting coordinates to
absolute screenshot pixels. For point annotations, the containing target
region serves as the reference box.

We also report action-type accuracy against the task label and exact agreement
with the original MNN output. These metrics ignore coordinate and argument
differences, which are captured by IoU and task-level success; together, they
test whether either transformation changes the selected operation or its
target.
Table~\ref{tab:quality-preservation} reports the resulting paired comparison.

\subsection{Scaling with Agent Steps}

\begin{figure}[ht]
\centering
\includegraphics[width=\columnwidth]{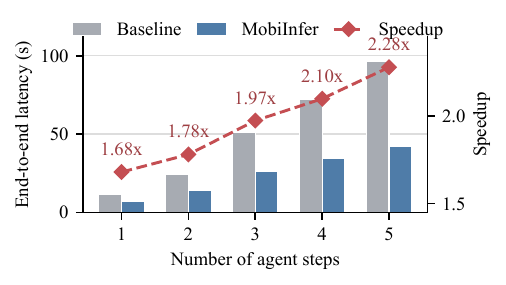}
\caption{End-to-end latency as the number of agent steps increases. Bars report
absolute latency for the baseline and full MobiInfer, while the dashed line and
right axis report the corresponding speedup.}
\label{fig:task-length-scaling}
\end{figure}

Figure~\ref{fig:task-length-scaling} evaluates whether MobiInfer's benefit
persists as task length grows across the full evaluation suite. From one to
five steps, baseline latency increases from 11.47 to 96.53 seconds, whereas
MobiInfer increases from 6.83 to 42.39 seconds. The resulting speedup rises
from 1.68$\times$ to 2.28$\times$, consistent with reusable prompt work and
structured decoding savings accumulating over repeated agent calls.

\begin{table}[ht]
\centering
\scriptsize
\setlength{\tabcolsep}{1.2pt}
\begin{tabular}{@{}lrrrr@{}}
\toprule
Variant & Mean IoU & IoU$\geq$0.5 & Action acc. & MNN agree. \\
\midrule
Orig. + full vocab. & 0.782 & 76.4\% & 91.8\% & 100.0\% \\
Canonical + full vocab. & 0.780 & 76.1\% & 91.7\% & 98.9\% \\
Orig. + pruned vocab. & 0.779 & 75.8\% & 91.5\% & 98.5\% \\
Canonical + pruned vocab. & 0.777 & 75.5\% & 91.3\% & 98.1\% \\
\bottomrule
\end{tabular}
\caption{Paired step-level quality comparison of full-vocabulary decoding and
MobiInfer's global-pruned, phase-indexed vocabulary path on identical
screenshots and task states.}
\label{tab:quality-preservation}
\end{table}

\subsection{Ablation Study}

\begin{figure}[ht]
\centering
\includegraphics[width=\columnwidth]{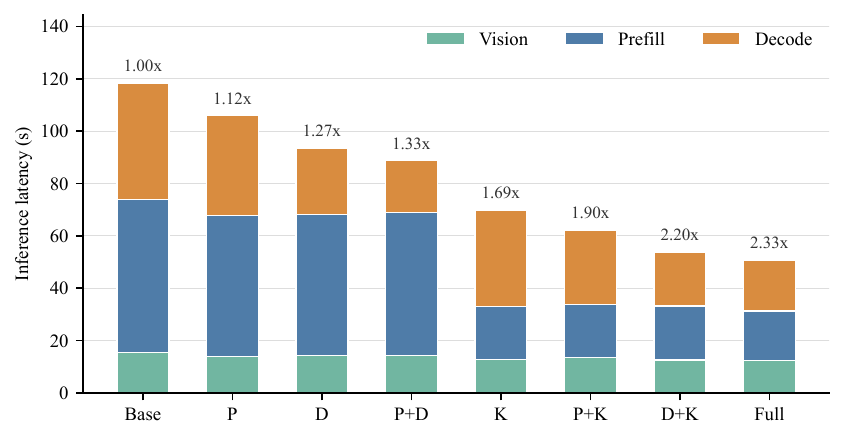}
\caption{Full-factorial component ablation on OnePlus 13 over the RealWorld dataset. Stacked bars report absolute inference latency in
seconds, decomposed into vision, prefill, and decode stages. The label above
each bar reports the corresponding end-to-end latency speedup relative to the
dense baseline. Every configuration has
three runs. P, K, and D denote phase-indexed vocabulary specialization, GUI
trajectory-aware prefill, and schema-guided drafting.}
\label{fig:component-ablation}
\end{figure}

Figure~\ref{fig:component-ablation} isolates the input- and output-side
mechanisms over the full evaluation suite of five-step tasks. The K variants
reduce the prefill portion of the stacked latency because they avoid
reprocessing the unchanged prompt prefix, while the D variants reduce the
decode portion by drafting only predictable output regions. The full
configuration reaches a $2.33\times$ end-to-end latency speedup in this
ablation suite. These bars report absolute stage latency; the stage-throughput
measurements in Table~\ref{tab:main-results} use the separate token-throughput
definition given in its caption.
The modest latency contribution of P indicates that its primary benefit is
reducing LM-head computation, memory, and deployment storage, while K and D
provide most of the end-to-end latency gain.

These results support the proposed lifetime- and phase-aware decomposition;
protecting action labels and coordinates preserves execution-critical behavior.

\subsection{Speculation Strategy Comparison}

We compare three speculation strategies under a common dense MNN execution
setting: MNN's native prompt-only lookahead, the learned EAGLE-3
decoder~\cite{li2025eagle3}, and MobiInfer's schema-guided drafting strategy.
The comparison isolates the effect of
the drafting mechanism while keeping the model, device, quantization, prompt
template, action schema, and task suite fixed. We report end-to-end latency,
decode throughput, and task success because a useful mobile speculation method
must improve latency without compromising action quality.

As a model-free baseline, we use MNN's native lookahead speculative decoder
~\cite{fu2024lookahead,jiang2020mnn} in
a prompt-only configuration. It retrieves repeated n-grams from the input
prompt, uses their continuations as drafts, and lets the target model verify
them in a batched forward pass; Draft additionally applies schema-aware phase
control and protected-field boundaries.

\begin{table}[ht]
\centering
\scriptsize
\setlength{\tabcolsep}{1pt}
\begin{tabular}{@{}lrrr@{}}
\toprule
Method & \shortstack{E2E latency\\(s)} &
\shortstack{Decode throughput\\speedup ($\times$)} &
\shortstack{Task success\\(\%)} \\
\midrule
MNN baseline & 118.3 & 1.00$\times$ & \textbf{42.3} \\
MNN + Lookahead & 103.4 & 1.23$\times$ & 41.4 \\
MNN + EAGLE-3 & 126.5 & 1.15$\times$ & 39.6 \\
MNN + \textbf{Schema-guided
drafting} & \textbf{93.2} & \textbf{1.84$\times$} & 41.4 \\
\bottomrule
\end{tabular}
\caption{Speculation strategy comparison for the MAI-UI-2B model on OnePlus 13
using the RealWorld dataset, with the same quantization, prompt
template, and action schema. }
\label{tab:eagle-comparison}
\end{table}

Table~\ref{tab:eagle-comparison} shows that EAGLE-3 improves decode throughput,
but its draft-model and verification
overheads increase end-to-end latency on mobile. Schema-guided drafting
improves both metrics without requiring a separately trained draft model while
preserving task success on this suite.

\subsection{Memory and Storage Footprint}

A substantial part of the exported mobile model is not in the transformer
blocks. In our exported MNN baseline, the token embedding and LM-head matrices
account for 34.3\% of the model weights. These matrices are often kept at higher
precision in mobile inference engines to preserve output quality, so they remain
a large memory and storage cost even after INT8 GPTQ-based post-training
quantization. MobiInfer
reduces this cost by pruning the embedding and LM-head matrices and using
phase-indexed heads: after pruning, their storage is reduced
to nearly one-tenth of the original size while preserving original token IDs and
byte-complete coverage.

Table~\ref{tab:resource-overhead} reports runtime memory and deployment storage.
MobiInfer reduces peak process PSS on every measured
platform, including a 31.82\% reduction on the OnePlus 13. Its auxiliary
artifacts do not erase the model-size saving: the full deployment remains
29.97\% smaller than the MNN baseline. No measured memory run uses swap.

\begin{table}[ht]
\centering
\scriptsize
\setlength{\tabcolsep}{2.5pt}
\begin{tabular}{@{}lrrrr@{}}
\toprule
\multicolumn{5}{@{}l@{}}{\textbf{Runtime memory (peak PSS, MiB)}} \\
\midrule
Platform & Baseline & MobiInfer & \multicolumn{2}{c}{Reduction} \\
\midrule
OnePlus 13 & 3769.98 & 2570.26 & \multicolumn{2}{c}{31.82\%} \\
OPPO Find X8 & 3765.24 & 2564.38 & \multicolumn{2}{c}{31.89\%} \\
Redmi K50 & 3728.04 & 2560.08 & \multicolumn{2}{c}{31.33\%} \\
\midrule
\multicolumn{5}{@{}l@{}}{\textbf{Deployment storage (MB)}} \\
\midrule
Deployment & Model & n-gram + KV & Total & Reduction \\
\midrule
MNN baseline & 3279.83 & -- & 3279.83 & -- \\
\textbf{MobiInfer} & \textbf{2161.25} & \textbf{135.75} & \textbf{2297.01} & \textbf{29.97\%} \\
\bottomrule
\end{tabular}
\caption{Runtime memory and deployment storage for the MAI-UI-2B model.
MobiInfer's auxiliary storage comprises a 27.48 MB n-gram table and a
108.27 MB cached-prompt KV set.}
\label{tab:resource-overhead}
\end{table}

%% file: sections/related_work.tex
\section{Related Work}

\paragraph{Mobile inference for GUI agents.}
On-device language and vision-language inference has mainly been advanced
through model compression, quantization, pruning, memory management, and
hardware-aware runtime optimization. MNN and llama.cpp provide general-purpose
inference runtimes for mobile and resource-constrained deployment
~\cite{jiang2020mnn,gerganov2023llamacpp}, while PowerInfer-2 and HeteroInfer study
smartphone LLM execution through hardware-aware inference and
scheduling~\cite{xue2024powerinfer2,chen2025heteroinfer}. Recent surveys
summarize the broader latency, memory, energy, and hardware-fragmentation
trade-offs of on-device LLM deployment
\cite{chen2026ondeviceSurvey,xu2024ondeviceSurvey}. In parallel,
mobile-GUI-agent work studies perception and grounding
~\cite{hong2024cogagent,you2024ferretui,cheng2024seeclick,wu2024osatlas,li2026goclick},
as well as planning, exploration, and task success
~\cite{zhang2023appagent,wang2024mobileagent,wen2024autodroid,nong2024mobileflow,lu2024guiodyssey,qin2025uitars,lin2025showui,zhang2025mobiagent,ye2025mobileagentv3,dai2025vdroid,xu2025mobilerl,wu2025verios,wang2026har,zhang2026omegause,tang2026expact,huang2026mobileexplorer,cai2026palmclaw}.
AITW and MobileAgentBench broaden mobile-agent evaluation
~\cite{rawles2023aitw,wang2024mobileagentbench}, while AndroidWorld provides
reproducible Android tasks and programmatic success
evaluators~\cite{rawles2025androidworld}. MobiInfer sits between these two
lines: it does not propose a new agent policy or a general compression method,
but accelerates the inference loop created when a multimodal GUI agent is run
fully on a phone. Recent models such as Ferret-UI Lite, LightAgent, and
AgentCPM-GUI target compact on-device agents, device--cloud execution, and
grounding-aware action generation, respectively
~\cite{yang2025ferretui,jiang2025lightagent,zhang2025agentcpm}; MobiInfer is
orthogonal, optimizing the repeated inference path for a fixed GUI-agent
workload.

\paragraph{Exposing reusable context.}
Prefix caching avoids recomputing attention states when requests share an
identical token prefix. Server-oriented systems provide the underlying cache
management and reuse mechanisms: vLLM's PagedAttention manages KV cache in
non-contiguous blocks~\cite{kwon2023vllm}, while SGLang reuses KV-cache prefixes
through RadixAttention~\cite{zheng2024sglang}. These systems rely on reuse
already exposed by
the request stream and require cached states to match the same token prefix.
MobiInfer instead canonicalizes GUI-agent prompts according to field lifetime,
making reusable context contiguous before cache restoration and delta prefill.

\paragraph{Accelerating structured action generation.}
Structured-generation systems such as SGLang, XGrammar, and XGrammar-2 provide
general abstractions for grammar-constrained decoding and structured LLM
program execution~\cite{zheng2024sglang,dong2025xgrammar,li2026xgrammar2};
INFERCEPT studies runtime support for augmented LLM
inference~\cite{abhyankar2024infercept}. Speculative decoding reduces
sequential target-model steps by verifying drafted tokens
\cite{leviathan2023speculative}, and learning-free n-gram speculation has been
shown to accelerate autoregressive inference~\cite{stewart2024ngrammys}.
Vocabulary selection and candidate screening further reduce large output-layer
costs~\cite{chen2019fastSoftmax,bian2026agentvocab}. MobiInfer binds these mechanisms to
GUI-agent output phases, which differ in projection cost, predictability, and
execution risk. It accelerates predictable structure and reasoning while
preserving target-model generation for execution-critical values.

%% file: sections/conclusion.tex
\hypertarget{conclusion}{%
\section{Conclusion}\label{conclusion}}

This paper presents \textbf{MobiInfer}, an inference engine for on-device
multimodal GUI agents. It exploits two properties of iterative agent inference:
prompt fields have different lifetimes, while output phases differ in
predictability and execution risk. Against the deployment MNN baseline, MobiInfer achieves
average speedups of $2.6\times$ in prefill, $2.5\times$ in decoding, and
$2.4\times$ in end-to-end task latency, while limiting task-success degradation
to less than one percentage point. It also reduces peak memory footprint and total deployment
storage by approximately 30\%. These results suggest that mobile agent inference
should be optimized around the repeated structure of the interaction loop,
rather than treated as independent generic model requests.

%% file: references.bib
@article{chen2026ondeviceSurvey,
  title = {On-device large language models: a survey of model compression and system optimization},
  author = {Chen, Wanyi and Wang, Junhao and Zhang, Yiwei and Shi, Yufan and Jiang, Tianyi and Zhou, Shengxian and Wu, Chenxu and Zhang, Andi and Zhou, Chenyue and Wang, Minxuan and Liu, Xinyu and Hao, Xiaoshuai and Wu, Yinan and Li, Yichen and Hu, Yuwei and Cao, Zhao and Lu, Yang and Li, Mengke and Ma, Yanbiao and Lu, Zhiwu and Han, Jungong and Guo, Yike},
  journal = {Artificial Intelligence Review},
  year = {2026},
  doi = {10.1007/s10462-026-11538-1}
}

@article{xu2024ondeviceSurvey,
  title = {On-Device Language Models: A Comprehensive Review},
  author = {Xu, Jiajun and Li, Zhiyuan and Chen, Wei and Wang, Qun and Gao, Xin and Cai, Qi and Ling, Ziyuan},
  journal = {arXiv preprint arXiv:2409.00088},
  year = {2024},
  doi = {10.48550/arXiv.2409.00088}
}

@inproceedings{kwon2023vllm,
  title = {Efficient Memory Management for Large Language Model Serving with {PagedAttention}},
  author = {Kwon, Woosuk and Li, Zhuohan and Zhuang, Siyuan and Sheng, Ying and Zheng, Lianmin and Yu, Cody Hao and Gonzalez, Joseph E. and Zhang, Hao and Stoica, Ion},
  booktitle = {Proceedings of the 29th Symposium on Operating Systems Principles},
  pages = {611--626},
  year = {2023},
  doi = {10.1145/3600006.3613165}
}

@inproceedings{leviathan2023speculative,
  title = {Fast Inference from Transformers via Speculative Decoding},
  author = {Leviathan, Yaniv and Kalman, Matan and Matias, Yossi},
  booktitle = {Proceedings of the 40th International Conference on Machine Learning},
  pages = {19274--19286},
  volume = {202},
  series = {Proceedings of Machine Learning Research},
  publisher = {PMLR},
  year = {2023},
  url = {https://proceedings.mlr.press/v202/leviathan23a.html}
}

@inproceedings{fu2024lookahead,
  title = {Break the Sequential Dependency of {LLM} Inference Using Lookahead Decoding},
  author = {Fu, Yichao and Bailis, Peter and Stoica, Ion and Zhang, Hao},
  booktitle = {Proceedings of the 41st International Conference on Machine Learning},
  pages = {14060--14079},
  year = {2024},
  volume = {235},
  publisher = {PMLR}
}

@inproceedings{chen2019fastSoftmax,
  title = {Learning to Screen for Fast Softmax Inference on Large Vocabulary Neural Networks},
  author = {Chen, Patrick H. and Si, Si and Li, Yang and Kumar, Sanjiv and Hsieh, Cho-Jui},
  booktitle = {International Conference on Learning Representations},
  year = {2019}
}

@inproceedings{bian2026agentvocab,
  title = {{AgentVocab}: Structure-Aware Vocabulary Adaptation for Efficient {LLM} Agents},
  author = {Bian, Kai and Mo, Haosi and Liu, Xuebo and Song, Shuangyong and Li, Jing and Li, Yongxiang and Zhang, Min and Li, Xuelong},
  booktitle = {Proceedings of the 43rd International Conference on Machine Learning},
  volume = {306},
  series = {Proceedings of Machine Learning Research},
  publisher = {PMLR},
  year = {2026}
}

@inproceedings{rawles2025androidworld,
  title = {{AndroidWorld}: A Dynamic Benchmarking Environment for Autonomous Agents},
  author = {Rawles, Christopher and Clinckemaillie, Sarah and Chang, Yifan and Waltz, Jonathan and Lau, Gabrielle and Fair, Marybeth and Li, Alice and Bishop, William E. and Li, Wei and Campbell-Ajala, Folawiyo and Toyama, Daniel and Berry, Robert and Tyamagundlu, Divya and Lillicrap, Timothy and Riva, Oriana},
  booktitle = {International Conference on Learning Representations},
  year = {2025},
  url = {https://openreview.net/forum?id=il5yUQsrjC}
}

@article{wang2024mobileagent,
  title = {Mobile-Agent: Autonomous Multi-Modal Mobile Device Agent with Visual Perception},
  author = {Wang, Junyang and Xu, Haiyang and Ye, Jiabo and Yan, Ming and Shen, Weizhou and Zhang, Ji and Huang, Fei and Sang, Jitao},
  journal = {arXiv preprint arXiv:2401.16158},
  year = {2024},
  doi = {10.48550/arXiv.2401.16158}
}

@inproceedings{wen2024autodroid,
  title = {{AutoDroid}: {LLM}-powered Task Automation in Android},
  author = {Wen, Hao and Li, Yuanchun and Liu, Guohong and Zhao, Shanhui and Yu, Tao and Li, Toby Jia-Jun and Jiang, Shiqi and Liu, Yunhao and Zhang, Yaqin and Liu, Yunxin},
  booktitle = {Proceedings of the 30th Annual International Conference on Mobile Computing and Networking},
  pages = {543--557},
  year = {2024},
  doi = {10.1145/3636534.3649379}
}

@article{nong2024mobileflow,
  title = {{MobileFlow}: A Multimodal {LLM} for Mobile {GUI} Agent},
  author = {Nong, Songqin and Zhu, Jiali and Wu, Rui and Jin, Jiongchao and Shan, Shuo and Huang, Xiutian and Xu, Wenhao},
  journal = {arXiv preprint arXiv:2407.04346},
  year = {2024},
  doi = {10.48550/arXiv.2407.04346}
}

@article{qin2025uitars,
  title = {{UI-TARS}: Pioneering Automated {GUI} Interaction with Native Agents},
  author = {Qin, Yujia and Ye, Yining and Fang, Junjie and Wang, Haoming and Liang, Shihao and Tian, Shizuo and Zhang, Junda and Li, Jiahao and Li, Yunxin and Huang, Shijue and Zhong, Wanjun and Li, Kuanye and Yang, Jiale and Miao, Yu and Lin, Woyu and Liu, Longxiang and Jiang, Xu and Ma, Qianli and Li, Jingyu and Xiao, Xiaojun and Cai, Kai and Li, Chuang and Zheng, Yaowei and Jin, Chaolin and Li, Chen and Zhou, Xiao and Wang, Minchao and Chen, Haoli and Li, Zhaojian and Yang, Haihua and Liu, Haifeng and Lin, Feng and Peng, Tao and Liu, Xin and Shi, Guang},
  journal = {arXiv preprint arXiv:2501.12326},
  year = {2025},
  doi = {10.48550/arXiv.2501.12326}
}

@article{dai2025vdroid,
  title = {Advancing Mobile {GUI} Agents: A Verifier-Driven Approach to Practical Deployment},
  author = {Dai, Gaole and Jiang, Shiqi and Cao, Ting and Li, Yuanchun and Yang, Yuqing and Tan, Rui and Li, Mo and Qiu, Lili},
  journal = {arXiv preprint arXiv:2503.15937},
  year = {2025},
  doi = {10.48550/arXiv.2503.15937}
}

@inproceedings{wang2026har,
  title = {History-Aware Reasoning for {GUI} Agents},
  author = {Wang, Ziwei and Yang, Leyang and Tang, Xiaoxuan and Zhou, Sheng and Chen, Dajun and Jiang, Wei and Li, Yong},
  booktitle = {Proceedings of the AAAI Conference on Artificial Intelligence},
  volume = {40},
  number = {43},
  pages = {36448--36456},
  year = {2026},
  doi = {10.1609/aaai.v40i43.40966}
}

@article{li2026goclick,
  title = {{GoClick}: Lightweight Element Grounding Model for Autonomous {GUI} Interaction},
  author = {Li, Hongxin and Chen, Yuntao and Zhang, Zhaoxiang},
  journal = {arXiv preprint arXiv:2604.23941},
  year = {2026},
  doi = {10.48550/arXiv.2604.23941}
}

@article{zhang2023appagent,
  title = {{AppAgent}: Multimodal Agents as Smartphone Users},
  author = {Zhang, Chi and Yang, Zhao and Liu, Jiaxuan and Han, Yucheng and Chen, Xin and Huang, Zebiao and Fu, Bin and Yu, Gang},
  journal = {arXiv preprint arXiv:2312.13771},
  year = {2023},
  doi = {10.48550/arXiv.2312.13771}
}

@inproceedings{hong2024cogagent,
  title = {{CogAgent}: A Visual Language Model for {GUI} Agents},
  author = {Hong, Wenyi and Wang, Weihan and Lv, Qingsong and Xu, Jiazheng and Yu, Wenmeng and Ji, Junhui and Wang, Yan and Wang, Zihan and Dong, Yuxiao and Ding, Ming and Tang, Jie},
  booktitle = {Proceedings of the IEEE/CVF Conference on Computer Vision and Pattern Recognition},
  pages = {14281--14290},
  year = {2024}
}

@inproceedings{you2024ferretui,
  title = {{Ferret-UI}: Grounded Mobile {UI} Understanding with Multimodal {LLM}s},
  author = {You, Keen and Zhang, Haotian and Schoop, Eldon and Weers, Floris and Swearngin, Amanda and Nichols, Jeffrey and Yang, Yinfei and Gan, Zhe},
  booktitle = {European Conference on Computer Vision},
  pages = {240--255},
  year = {2024},
  doi = {10.1007/978-3-031-73039-9_14}
}

@inproceedings{cheng2024seeclick,
  title = {{SeeClick}: Harnessing {GUI} Grounding for Advanced Visual {GUI} Agents},
  author = {Cheng, Kanzhi and Sun, Qiushi and Chu, Yougang and Xu, Fangzhi and Li, Yantao and Zhang, Jianbing and Wu, Zhiyong},
  booktitle = {Proceedings of the 62nd Annual Meeting of the Association for Computational Linguistics (Volume 1: Long Papers)},
  pages = {9313--9332},
  publisher = {Association for Computational Linguistics},
  year = {2024},
  doi = {10.18653/v1/2024.acl-long.505}
}

@article{lu2024guiodyssey,
  title = {{GUI} Odyssey: A Comprehensive Dataset for Cross-App {GUI} Navigation on Mobile Devices},
  author = {Lu, Quanfeng and Shao, Wenqi and Liu, Zitao and Meng, Fanqing and Li, Boxuan and Chen, Botong and Huang, Siyuan and Zhang, Kaipeng and Qiao, Yu and Luo, Ping},
  journal = {arXiv preprint arXiv:2406.08451},
  year = {2024},
  doi = {10.48550/arXiv.2406.08451}
}

@article{wu2024osatlas,
  title = {{OS-ATLAS}: A Foundation Action Model for Generalist {GUI} Agents},
  author = {Wu, Zhiyong and Wu, Zhenyu and Xu, Fangzhi and Wang, Yian and Sun, Qiushi and Jia, Chengyou and Cheng, Kanzhi and Ding, Zichen and Chen, Liheng and Liang, Paul Pu and Qiao, Yu},
  journal = {arXiv preprint arXiv:2410.23218},
  year = {2024},
  doi = {10.48550/arXiv.2410.23218}
}

@inproceedings{lin2025showui,
  title = {{ShowUI}: One Vision-Language-Action Model for {GUI} Visual Agent},
  author = {Lin, Kevin Qinghong and Li, Linjie and Gao, Difei and Yang, Zhengyuan and Wu, Shiwei and Bai, Zechen and Lei, Weixian and Wang, Lijuan and Shou, Mike Zheng},
  booktitle = {Proceedings of the IEEE/CVF Conference on Computer Vision and Pattern Recognition},
  year = {2025}
}

@article{xu2025mobilerl,
  title = {{MobileRL}: Online Agentic Reinforcement Learning for Mobile {GUI} Agents},
  author = {Xu, Yifan and Liu, Xiao and Liu, Xinghan and Fu, Jiaqi and Zhang, Hanchen and Jing, Bohao and Zhang, Shudan and Wang, Yuting and Zhao, Wenyi and Dong, Yuxiao},
  journal = {arXiv preprint arXiv:2509.18119},
  year = {2025},
  doi = {10.48550/arXiv.2509.18119}
}

@article{wu2025verios,
  title = {{VeriOS}: Query-Driven Proactive Human-Agent-{GUI} Interaction for Trustworthy {OS} Agents},
  author = {Wu, Zheng and Huang, Heyuan and Lou, Xingyu and Qu, Xiangmou and Cheng, Pengzhou and Wu, Zongru and Liu, Weiwen and Zhang, Weinan and Wang, Jun and Wang, Zhaoxiang and Zhang, Zhuosheng},
  journal = {arXiv preprint arXiv:2509.07553},
  year = {2025},
  doi = {10.48550/arXiv.2509.07553}
}

@article{zhang2026omegause,
  title = {{OmegaUse}: Building a General-Purpose {GUI} Agent for Autonomous Task Execution},
  author = {Zhang, Le and Xiao, Yixiong and Lu, Xinjiang and Cao, Jingjia and Zhao, Yusai and Zhou, Jingbo and An, Lang and Feng, Zikan and Sha, Wanxiang and Shi, Yu and Xiao, Congxi and Xiong, Jian and Zhang, Yankai and Wu, Hua and Wang, Haifeng},
  journal = {arXiv preprint arXiv:2601.20380},
  year = {2026},
  doi = {10.48550/arXiv.2601.20380}
}

@misc{tang2026expact,
  title = {{ExpAct}: Building and Evolving Structural and Procedural Experience for Mobile {GUI} Agents},
  author = {Tang, Shunye and Chai, Yuxiang and Guo, Yaxuan and Han, Rongduo and Lan, Mingang and Liu, Liang and Liu, Guangyi and Zhao, Pengxiang and Xiao, Han and Luo, Qinyi and Wang, Guozhi and Wang, Hao and Liang, Xiao and Zhang, Haining and Wang, Wei},
  year = {2026},
  note = {Submitted to ICML 2026},
  url = {https://openreview.net/forum?id=qkSahzB9hh}
}

@inproceedings{rawles2023aitw,
  title = {{AndroidInTheWild}: A Large-Scale Dataset for Android Device Control},
  author = {Rawles, Christopher and Li, Alice and Rodriguez, Daniel and Riva, Oriana and Lillicrap, Timothy},
  booktitle = {Advances in Neural Information Processing Systems},
  volume = {36},
  year = {2023}
}

@article{wang2024mobileagentbench,
  title = {{MobileAgentBench}: An Efficient and User-Friendly Benchmark for Mobile {LLM} Agents},
  author = {Wang, Luyuan and Deng, Yongyu and Zha, Yiwei and Mao, Guodong and Wang, Qinmin and Min, Tianchen and Chen, Wei and Chen, Shoufa},
  journal = {arXiv preprint arXiv:2406.08184},
  year = {2024},
  doi = {10.48550/arXiv.2406.08184}
}

@article{zhang2025mobiagent,
  title = {{MobiAgent}: A Systematic Framework for Customizable Mobile Agents},
  author = {Zhang, Cheng and Feng, Erhu and Zhao, Xi and Zhao, Yisheng and Gong, Wangbo and Sun, Jiahui and Du, Dong and Hua, Zhichao and Xia, Yubin and Chen, Haibo},
  journal = {arXiv preprint arXiv:2509.00531},
  year = {2025}
}

@article{zhou2025maiui,
  title = {{MAI-UI} Technical Report: Real-World Centric Foundation {GUI} Agents},
  author = {Zhou, Hanzhang and Zhang, Xu and Tong, Panrong and Zhang, Jianan and Chen, Liangyu and Kong, Quyu and Cai, Chenglin and Liu, Chen and Wang, Yue and Zhou, Jingren and Hoi, Steven},
  journal = {arXiv preprint arXiv:2512.22047},
  year = {2025}
}

@misc{intel2023autoround,
  title = {{AutoRound}: A SOTA Quantization Algorithm for High-Accuracy Low-Bit {LLM} Inference},
  author = {{Intel}},
  year = {2023},
  howpublished = {\url{https://github.com/intel/auto-round}}
}

@misc{stepfun2025gelabzero,
  title = {{GELab-Zero-4B}},
  author = {{StepFun-GELab Team}},
  year = {2025},
  howpublished = {\url{https://huggingface.co/stepfun-ai/GELab-Zero-4B-preview}}
}

@inproceedings{frantar2023gptq,
  title = {{GPTQ}: Accurate Post-Training Quantization for Generative Pre-trained Transformers},
  author = {Frantar, Elias and Ashkboos, Saleh and Hoefler, Torsten and Alistarh, Dan},
  booktitle = {International Conference on Learning Representations},
  year = {2023},
  url = {https://openreview.net/forum?id=tcbBPnfwxS}
}

@article{ye2025mobileagentv3,
  title = {Mobile-Agent-v3: Fundamental Agents for GUI Automation},
  author = {Ye, Jiabo and Zhang, Xi and Xu, Haiyang and Liu, Haowei and Wang, Junyang and Zhu, Zhaoqing and Zheng, Ziwei and Gao, Feiyu and Cao, Junjie and Lu, Zhengxi and Liao, Jitong and Zheng, Qi and Huang, Fei and Zhou, Jingren and Yan, Ming},
  journal = {arXiv preprint arXiv:2508.15144},
  year = {2025},
  doi = {10.48550/arXiv.2508.15144}
}

@inproceedings{jiang2020mnn,
  title = {{MNN}: A Universal and Efficient Inference Engine},
  author = {Jiang, Xiaotang and Wang, Huan and Chen, Yiliu and Wu, Ziqi and Wang, Lichuan and Zou, Bin and Yang, Yafeng and Cui, Zongyang and Cai, Yu and Yu, Tianhang and Lv, Chengfei and Wu, Zhihua},
  booktitle = {Proceedings of Machine Learning and Systems},
  year = {2020}
}

@misc{gerganov2023llamacpp,
  title = {{llama.cpp}: {LLM} Inference in {C}/{C}++},
  author = {Gerganov, Georgi and contributors},
  year = {2023},
  howpublished = {\url{https://github.com/ggml-org/llama.cpp}}
}

@inproceedings{zheng2024sglang,
  title = {{SGLang}: Efficient Execution of Structured Language Model Programs},
  author = {Zheng, Lianmin and Yin, Liangsheng and Xie, Zhiqiang and Sun, Chuyue and Huang, Jeff and Yu, Cody Hao and Cao, Shiyi and Kozyrakis, Christos and Stoica, Ion and Gonzalez, Joseph E. and Barrett, Clark and Sheng, Ying},
  booktitle = {Advances in Neural Information Processing Systems},
  year = {2024}
}

@inproceedings{dong2025xgrammar,
  title = {{XGrammar}: Flexible and Efficient Structured Generation Engine for Large Language Models},
  author = {Dong, Yixin and Ruan, Charlie F. and Cai, Yaxing and Lai, Ruihang and Xu, Ziyi and Zhao, Yilong and Chen, Tianqi},
  booktitle = {Proceedings of Machine Learning and Systems},
  year = {2025}
}

@article{li2026xgrammar2,
  title = {{XGrammar-2}: Efficient Dynamic Structured Generation Engine for Agentic {LLM}s},
  author = {Li, Linzhang and Dong, Yixin and Wang, Guanjie and Xu, Ziyi and Jiang, Alexander and Chen, Tianqi},
  journal = {arXiv preprint arXiv:2601.04426},
  year = {2026}
}

@inproceedings{abhyankar2024infercept,
  title = {{INFERCEPT}: Efficient Intercept Support for Augmented Large Language Model Inference},
  author = {Abhyankar, Reyna and He, Zijian and Srivatsa, Vikranth and Zhang, Hao and Zhang, Yiying},
  booktitle = {Proceedings of the 41st International Conference on Machine Learning},
  pages = {81--95},
  volume = {235},
  series = {Proceedings of Machine Learning Research},
  publisher = {PMLR},
  year = {2024},
  url = {https://proceedings.mlr.press/v235/abhyankar24a.html}
}

@article{xue2024powerinfer2,
  title = {{PowerInfer-2}: Fast Large Language Model Inference on a Smartphone},
  author = {Xue, Zhenliang and Song, Yixin and Mi, Zeyu and Zheng, Xinrui and Xia, Yubin and Chen, Haibo},
  journal = {arXiv preprint arXiv:2406.06282},
  year = {2024}
}

@inproceedings{chen2025heteroinfer,
  title = {Characterizing Mobile {SoC} for Accelerating Heterogeneous {LLM} Inference},
  author = {Chen, Le and Feng, Dahu and Feng, Erhu and Wang, Yingrui and Zhao, Rong and Xia, Yubin and Xu, Pinjie and Chen, Haibo},
  booktitle = {Proceedings of the 31st Symposium on Operating Systems Principles},
  pages = {359--374},
  year = {2025},
  doi = {10.1145/3731569.3764808}
}

@article{huang2026mobileexplorer,
  title = {{MobileExplorer}: Accelerating On-Device Inference for Mobile {GUI} Agents via Online Exploration},
  author = {Huang, Runxi and Zhang, Liyu and Liu, Shengzhong and Ouyang, Xiaomin},
  journal = {arXiv preprint arXiv:2605.26546},
  year = {2026}
}

@article{cai2026palmclaw,
  title = {{PalmClaw}: A Native On-Device Agent Framework for Mobile Phones},
  author = {Cai, Hongru and Li, Yongqi and Wei, Ran and Li, Wenjie},
  journal = {arXiv preprint arXiv:2607.13027},
  year = {2026}
}

@inproceedings{stewart2024ngrammys,
  title = {The {N-Grammys}: Accelerating Autoregressive Inference with Learning-Free Batched Speculation},
  author = {Stewart, Lawrence and Trager, Matthew and Gonugondla, Sujan and Soatto, Stefano},
  booktitle = {Proceedings of the 4th NeurIPS Workshop on Efficient Natural Language and Speech Processing},
  year = {2024}
}

@inproceedings{li2025eagle3,
  title = {{EAGLE-3}: Scaling up Inference Acceleration of Large Language Models via Training-Time Test},
  author = {Li, Yuhui and Wei, Fangyun and Zhang, Chao and Zhang, Hongyang},
  booktitle = {Advances in Neural Information Processing Systems},
  year = {2025},
  note = {NeurIPS 2025},
  eprint = {2503.01840},
  archivePrefix = {arXiv}
}

@article{yang2025ferretui,
  title = {Ferret-UI Lite: Lessons from Building Small On-Device GUI Agents},
  author = {Yang, Zhen and Dou, Zi-Yi and Feng, Di and Huang, Forrest and Nguyen, Anh and You, Keen and Attia, Omar and Yang, Yuhao and Feng, Michael and Zhang, Haotian and Ramrakhya, Ram and Jia, Chao and Nichols, Jeffrey and Toshev, Alexander and Yang, Yinfei and Gan, Zhe},
  journal = {arXiv preprint arXiv:2509.26539},
  year = {2025}
}

@article{jiang2025lightagent,
  title = {LightAgent: Mobile Agentic Foundation Models},
  author = {Jiang, Yangqin and Huang, Chao},
  journal = {arXiv preprint arXiv:2510.22009},
  year = {2025}
}

@inproceedings{zhang2025agentcpm,
  title={Agentcpm-gui: Building mobile-use agents with reinforcement fine-tuning},
  author={Zhang, Zhong and Lu, Yaxi and Fu, Yikun and Huo, Yupeng and Yang, Shenzhi and Wu, Yesai and Si, Han and Cong, Xin and Chen, Haotian and Lin, Yankai and others},
  booktitle={Proceedings of the 2025 Conference on Empirical Methods in Natural Language Processing: System Demonstrations},
  pages={155--180},
  year={2025}
}
